\documentclass[structabstract]{aa}  
\usepackage{mathrsfs}
\usepackage{hyperref}
\usepackage{graphicx}
\usepackage{epstopdf}
\usepackage{txfonts}
\usepackage{natbib}
\bibpunct{(}{)}{;}{a}{}{,} % to follow the A&A style
\begin{document}
   \title{Asteroseismology from multi-month \emph{Kepler} photometry:\\
   the evolved Sun-like stars KIC~10273246 and KIC~10920273}

   %\subtitle{}

   \author{
   T.~L.~Campante\inst{\ref{inst1},\ref{inst2}}\fnmsep\thanks{\email{campante@astro.up.pt,campante@phys.au.dk}}
   \and R.~Handberg\inst{\ref{inst2}}
   \and S.~Mathur\inst{\ref{inst3}}
   \and T.~Appourchaux\inst{\ref{inst5}}
   \and T.~R.~Bedding\inst{\ref{inst6}}
   \and W.~J.~Chaplin\inst{\ref{inst7}}
   \and R.~A.~Garc\'ia\inst{\ref{inst4}}
   \and \\ B.~Mosser\inst{\ref{inst13}}
   \and O.~Benomar\inst{\ref{inst5}}
   \and A.~Bonanno\inst{\ref{inst14}}
   \and E.~Corsaro\inst{\ref{inst14}}
   \and S.~T.~Fletcher\inst{\ref{inst11}}   
   \and P.~Gaulme\inst{\ref{inst5}}
   \and S.~Hekker\inst{\ref{inst7},\ref{inst19}}
   \and C.~Karoff\inst{\ref{inst2}}
   \and \\ C.~R\'egulo\inst{\ref{inst8},\ref{inst9}}
   \and D.~Salabert\inst{\ref{inst8},\ref{inst9}}
   \and G.~A.~Verner\inst{\ref{inst7},\ref{inst10}}
   \and T.~R.~White\inst{\ref{inst6},\ref{inst16}}
   \and G.~Houdek\inst{\ref{inst12}}
   \and I.~M.~Brand\~ao\inst{\ref{inst1}}
   \and O.~L.~Creevey\inst{\ref{inst8},\ref{inst9}}
   \and \\ G.~Do\u{g}an\inst{\ref{inst2}}
   \and M.~Bazot\inst{\ref{inst1}}
   \and J.~Christensen-Dalsgaard\inst{\ref{inst2}}
   \and M.~S.~Cunha\inst{\ref{inst1}}
   \and Y.~Elsworth\inst{\ref{inst7}}
   \and D.~Huber\inst{\ref{inst6}}
   \and H.~Kjeldsen\inst{\ref{inst2}}
   \and \\ M.~Lundkvist\inst{\ref{inst2}}
   \and J.~Molenda-\.Zakowicz\inst{\ref{inst15}}
   \and M.~J.~P.~F.~G.~Monteiro\inst{\ref{inst1}}
   \and D.~Stello\inst{\ref{inst6}}
   \and \\ B.~D.~Clarke\inst{\ref{inst17}}
   \and F.~R.~Girouard\inst{\ref{inst18}}
   \and J.~R.~Hall\inst{\ref{inst18}}
          }

   \institute{
    Centro de Astrof\'isica, DFA-Faculdade de Ci\^encias, Universidade do Porto, Rua das Estrelas, 4150-762 Porto, Portugal\label{inst1}
    \and Department of Physics and Astronomy, Aarhus University, DK-8000 Aarhus C, Denmark\label{inst2}
    \and High Altitude Observatory, National Center for Atmospheric Research, Boulder, Colorado 80307, USA\label{inst3}
    \and Institut d'Astrophysique Spatiale, Universit\'e Paris XI -- CNRS (UMR8617), Batiment 121, 91405 Orsay Cedex, France\label{inst5}
    \and Sydney Institute for Astronomy (SIfA), School of Physics, University of Sydney, NSW 2006, Australia\label{inst6}
    \and School of Physics and Astronomy, University of Birmingham, Edgbaston, Birmingham, B15 2TT, UK\label{inst7}
    \and Laboratoire AIM, CEA/DSM-CNRS-Universit\'e Paris Diderot; IRFU/SAp, Centre de Saclay, 91191, Gif-sur-Yvette, France\label{inst4}
    \and LESIA, CNRS, Universit\'e Pierre et Marie Curie, Universit\'e Paris Diderot, Observatoire de Paris, 92195 Meudon Cedex, France\label{inst13}
    \and INAF Osservatorio Astrofisico di Catania, Via S.~Sofia 78, 95123, Catania, Italy\label{inst14}
    \and Materials Engineering Research Institute, Faculty of Arts, Computing, Engineering and Sciences, Sheffield Hallam University, Sheffield, S1 1WB, UK\label{inst11}
    \and Astronomical Institute Anton Pannekoek, University of Amsterdam, Science Park 904, 1098 XH Amsterdam, The Netherlands\label{inst19}
    \and Departamento de Astrof\'isica, Universidad de La Laguna, E-38206 La Laguna, Tenerife, Spain\label{inst8}
    \and Instituto de Astrof\'isica de Canarias, E-38200 La Laguna, Tenerife, Spain\label{inst9}
    \and Astronomy Unit, Queen Mary, University of London, Mile End Road, London, E1 4NS, UK\label{inst10}
    \and Australian Astronomical Observatory, PO Box 296, Epping NSW 1710, Australia\label{inst16}
    \and Institute of Astronomy, University of Vienna, A-1180 Vienna, Austria\label{inst12}
    \and Astronomical Institute, University of Wroc\l{}aw, ul.~Kopernika 11, 51-622 Wroc\l{}aw, Poland\label{inst15}
    \and SETI Institute/NASA Ames Research Center, Moffett Field, CA 94035, USA\label{inst17}
    \and Orbital Sciences Corporation/NASA Ames Research Center, Moffett Field, CA 94035, USA\label{inst18}
                 }

   %\date{Received September 15, 1996; accepted March 16, 1997}

% \abstract{}{}{}{}{} 
% 5 {} token are mandatory
 
  \abstract
   {The evolved main-sequence Sun-like stars KIC~10273246 (F-type) and KIC~10920273 (G-type) were observed with the NASA \emph{Kepler} satellite for approximately ten months with a duty cycle in excess of 90\%. Such continuous and long observations are unprecedented for solar-type stars other than the Sun.}
   {We aimed mainly at extracting estimates of p-mode frequencies -- as well as of other individual mode parameters -- from the power spectra of the light curves of both stars, thus providing scope for a full seismic characterization.}
   {The light curves were corrected for instrumental effects in a manner independent of the \emph{Kepler} Science Pipeline. Estimation of individual mode parameters was based both on the maximization of the likelihood of a model describing the power spectrum and on a classic prewhitening method. Finally, we employed a procedure for selecting frequency lists to be used in stellar modeling.}
   {A total of 30 and 21 modes of degree $l\!=\!0,1,2$ -- spanning at least eight radial orders -- have been identified for KIC~10273246 and KIC~10920273, respectively. Two avoided crossings ($l\!=\!1$ ridge) have been identified for KIC~10273246, whereas one avoided crossing plus another likely one have been identified for KIC~10920273. Good agreement is found between observed and predicted mode amplitudes for the F-type star KIC~10273246, based on a revised scaling relation. Estimates are given of the rotational periods, the parameters describing stellar granulation and the global asteroseismic parameters $\Delta\nu$ and $\nu_{\rm{max}}$.}
   {}


   \titlerunning{Asteroseismology from multi-month \emph{Kepler} photometry}
   \authorrunning{T. L. Campante et al.}

   \maketitle

\section{Introduction}

The NASA \emph{Kepler} Mission was designed to use the transit method to detect Earth-like planets in and near the habitable zones of late-type main-sequence stars \citep{Kp_planets,Kp_design}. The satellite consists of a 0.95-meter aperture telescope with a CCD array and is capable of producing photometric observations with a precision of a few parts-per-million (ppm) during a period of 4--6 years. The high-quality data provided by \emph{Kepler} are also well suited for conducting asteroseismic studies of stars as part of the \emph{Kepler} Asteroseismic Investigation \citep[KAI;][]{KAI}. Photometry of the vast majority of these stars is conducted at long cadence (29.4 minutes), while a revolving selection of up to 512 stars are monitored at short cadence (58.85 seconds). Short-cadence data allow us to investigate solar-like oscillations in main-sequence stars and subgiants, whose dominant periods are of the order of several minutes \citep{Kp_Chaplin,Kp_CD,Gemma}.  

The information contained in solar-like oscillations allows fundamental stellar properties (e.g., mass, radius and age) to be determined \citep[e.g.,][]{Stello1,Kp_CD,KallingerRG}. The internal stellar structure can be constrained to unprecedented levels, provided that individual oscillation mode parameters are extracted \citep[e.g.,][]{Cunha07}. This is possible in the case of the highest signal-to-noise ratio (SNR) observations, leading us to hope that asteroseismology will produce significant improvement on the theories related to stellar structure and evolution, on topics as diverse as energy generation and transport, rotation and stellar cycles \citep[e.g.,][]{Karoff09,Rafa10}.

Solar-like oscillations in a few tens of main-sequence stars and subgiants have been previously measured using ground-based high-precision spectroscopy \citep[e.g.,][]{Procyon} and the space-based photometric mission \emph{CoRoT} \citep[e.g.,][]{Michel08}. During the first seven months of \emph{Kepler} science operations, an asteroseismic survey of solar-type stars made it possible to detect solar-like oscillations in about 500 targets \citep{ChaplinScience}. This constitutes an increase of one order of magnitude in the number of such stars with confirmed oscillations. This large, homogeneous data sample opens the possibility of conducting ensemble asteroseismology.

Since the start of \emph{Kepler} science operations in 2009 May, a selection of solar-type stars have been continuously monitored at short cadence for more than seven months in order to test and validate the time-series photometry. Such continuous and long observations, also previously achieved by \emph{CoRoT} \citep[e.g.,][]{BBC}, are unprecedented for solar-type stars other than the Sun. We present herein the analysis of two of these stars{\footnote{Within the \emph{Kepler} Asteroseismic Science Consortium (KASC), KIC~10273246 is referred to as ``Mulder" and KIC~10920273 as ``Scully".}}, namely, \object{KIC~10273246} and \object{KIC~10920273}, both displaying relatively low SNR in the p-mode peaks. Two other stars, namely, KIC~11395018 and KIC~11234888, are analysed in a companion paper \citep{FurryMathur}. The analysis of these four stars has been conducted in a way so as to group together stars observed for the same length of time.

The two solar-type stars selected for this study are relatively faint (see Table \ref{KIC}) if we bear in mind that the apparent magnitude target range for detection of solar-like oscillations with \emph{Kepler} spans $Kp\!\approx\!6.5$ to $Kp\!\approx\!12.5$ \citep[see fig.~5 of][]{Chaplinpredict}. The \emph{Kepler} Input Catalog\footnote{\url{http://archive.stsci.edu/kepler/kepler\_fov/search.php}} \citep[KIC; e.g.,][]{KIC,Batalha,KIC2}, from which all KASC targets have been selected, classifies KIC~10273246 as an F-type star and KIC~10920273 as a G-type star (Table \ref{KIC}). The atmospheric parameters provided by the KIC -- as derived from photometric observations acquired in the Sloan filters -- do not have sufficient precision for asteroseismology. Although we can apply scaling relations to convert the KIC parameters of these targets into predicted seismic and non-resonant background parameters, as well as fundamental stellar properties, caution is needed if use is to be made of these derived quantities. Tighter constraints on $T_{\rm{eff}}$, $\log g$ and [\element{Fe}/\element{H}] will be obtained from spectra collected for these two targets with the FIES spectrograph at the Nordic Optical Telescope (Creevey et al., in preparation).
\begin{table*}[!t]
\caption{\label{KIC}Information given in the \emph{Kepler} Input Catalog.}
\centering
\begin{tabular}{lccccccc}
\hline\hline
Star&2MASS ID&$Kp$&$T_{\rm{eff}}$&Recalibrated\tablefootmark{a} $T_{\rm{eff}}$&$\log g$&[\element{Fe}/\element{H}]&$R$\\
&&&(K)&(K)&(dex)&(dex)&($R_{\sun}$)\\
\hline
KIC~10273246&19260576+4721300&10.90&$6074\pm200$&$6380\pm76$&$4.2\pm0.5$&$-0.3\pm0.5$&1.506\\
KIC10920273&19274576+4819454&11.93&$5574\pm200$&$5880\pm53$&$4.1\pm0.5$&$-0.4\pm0.5$&1.594\\
\hline
\end{tabular}
\tablefoot{\\
\tablefoottext{a}{A recalibration of the KIC photometry has posteriorly been performed by Pinsonneault et al.~(in preparation).}
}
\end{table*}

The outline of the paper is as follows: We start in Sect.~\ref{theory} by providing some background information on the properties of solar-like oscillations. In Sect.~\ref{strategy} we give an overview of the different peak-bagging\footnote{The term \emph{peak-bagging} refers to the extraction of individual mode parameters from the power spectrum of a light curve.} strategies employed and define a recipe for selecting frequency lists. Section \ref{analysis} is devoted to a thorough analysis of the power spectra of the time series. A summary and conclusions are presented in Sect.~\ref{conclusions}.

\section{Properties of solar-like oscillations}\label{theory}
Solar-like oscillations are predominantly global standing acoustic waves. These are p modes (pressure playing the role of the restoring force) and are characterized by being intrinsically damped while simultaneously stochastically excited by near-surface convection \citep[e.g.,][]{CD04}. Therefore, all stars cool enough to harbor an outer convective envelope -- whose locus in the H-R diagram approximately extends from the cool edge of the Cepheid instability strip and includes the red giant branch -- may be expected to exhibit solar-like oscillations.

Modes of oscillation are characterized by three quantum numbers: $n$, $l$ and $m$. The radial order $n$ characterizes the behavior of the mode in the radial direction. The degree $l$ and the azimuthal order $m$ determine the spherical harmonic describing the properties of the mode as a function of colatitude and longitude. In the case of stellar observations, the associated whole-disk light integration and consequent lack of spatial resolution strongly suppress the signal from all but the modes of the lowest degree (with $l\!\leq\!3$). For a spherically symmetric non-rotating star, mode frequencies depend only on $n$ and $l$. 

The observed modes of oscillation are typically high-order acoustic modes. If interaction with a g-mode (gravity playing the role of the restoring force) can be neglected, linear, adiabatic, high-order acoustic modes, in a spherically symmetric star, satisfy an asymptotic relation for the frequencies \citep{Vandakurov,Tassoul}:
	\begin{equation}\label{asymptotic}
		\nu_{nl}\!\sim\!\Delta\nu\,(n+l/2+\varepsilon)-l(l+1)D_0 \, ,
	\end{equation}
where the large frequency separation $\Delta \nu$ is the inverse of the sound travel time across the stellar diameter, $\varepsilon$ is a phase mostly sensitive to the properties of the near-surface region, and $D_0$ is a parameter sensitive to the sound-speed gradient near the core. The regular spacing of the frequency spectrum as conveyed by Eq.~\ref{asymptotic} is a characteristic feature of solar-like oscillations. We should, however, bear in mind that Eq.~\ref{asymptotic} is only an approximation. The large frequency separation does in fact depend both on frequency and on mode degree, being defined as
	\begin{equation}
		\Delta\nu_{nl}=\nu_{nl}-\nu_{n-1\,l}\!\approx\!\Delta\nu \, .
	\end{equation}
It is also conventional to define a so-called small frequency separation, also varying with frequency:
	\begin{equation}
		\delta\nu_{nl}=\nu_{nl}-\nu_{n-1\,l+2}\!\approx\!(4l+6)D_0 \, .
	\end{equation}
The large frequency separation essentially scales with the square root of the mean stellar density \citep[e.g.,][]{BrownGilliland}. Furthermore, the small frequency separation is sensitive to the structure of the core, decreasing with increasing stellar age. These two quantities thus have great diagnostic potential \citep[e.g.,][]{CD93,Deheuvels}.

Sharp variations in the stellar interior cause detectable oscillatory signals in the frequencies, also visible in the behavior of $\Delta \nu$ as a function of frequency \citep[e.g.,][]{Monteiro2000,Ballot04,Basu04,CuMe07,Houdek07}. These sharp features are mainly linked to borders of convection zones and to regions of rapid variation in the sound speed due to ionization of a dominant element. Their combined signature is detectable in frequencies of low-degree modes and such an analysis becomes possible in the stellar case once frequency precision is sufficiently high.

Stellar rotation, as well as any other physical process resulting in departure from spherical symmetry, introduces a dependence of the frequencies of non-radial modes on $m$. When the cyclic rotational frequency of the star, $\nu_{\rm{rot}}$, is small and in the case of rigid-body rotation dominated by advection, the cyclic frequency of a non-radial mode is given to first order by \citep{Ledoux}:
	\begin{equation}
		\label{ledoux}
		\nu_{nl m}=\nu_{nl0} + m\,\nu_{\rm{rot}} \, , \;\; |m| \leq l \, .
	\end{equation}
To a second order of approximation, centrifugal effects that disrupt the equilibrium structure of the star are taken into account through an additional frequency perturbation (independent of the sign of $m$). This perturbation in turn scales as the ratio of the centrifugal to the gravitational forces at the stellar surface, i.e., $\Omega^2R^3/(GM)$, where $\Omega$ denotes the surface angular velocity, $R$ the radius of the star, $M$ its mass, and $G$ the universal gravitational constant. Although negligible in the Sun, these effects may be significant for faster-rotating solar-type stars \citep[e.g.,][]{Ballotrot}. Large-scale magnetic fields may also introduce further corrections to the oscillation frequencies.

The frequency dependence of the mode surface amplitudes is determined both by (i) the frequency dependence of the stochastic process of excitation (mode energies result from a balance between the frequency-dependent energy input and the damping rate) and by (ii) the mode properties in the region of vigorous convection \citep[e.g.,][]{Houdek99,Samadi07}. The stochastic process of excitation is characterized by a relatively slow variation with frequency, meaning that it excites modes over a large frequency interval to comparable surface amplitudes. At low frequencies modes are evanescent in the region of efficient excitation, leading to small surface amplitudes. At high frequencies -- greater than or equal to the acoustic cut-off frequency -- modes undergo considerable energy loss through running waves in the atmosphere. Excitation is most efficient for those modes whose periods match the timescale of the near-surface convection. Also, the frequency of maximum amplitude, $\nu_{\rm{max}}$, is supposed to scale with the acoustic cut-off frequency, $\nu_{\rm{ac}}$ \citep{Brown,KB95}. All this gives rise to a characteristic distribution of power with frequency which is a signature of the presence of solar-like oscillations.

Substantial changes in the properties of solar-like oscillations occur with stellar evolution, particularly following the exhaustion of hydrogen in the core. Most noticeable is the occurrence of avoided crossings due to coupling between p and g modes of like degree \citep{Osaki,Aizenman}, which lead to significant departures from the regular frequency spacing described by Eq.~\ref{asymptotic} in the case of evolved stars. The frequencies of non-radial modes, in particular those of $l\!=\!1$ modes, are shifted by avoided crossings when they couple with g modes trapped in the deep stellar interior. At the avoided crossings these modes have a mixed nature, with both p- and g-mode behavior. Provided they are excited to observable amplitudes (their high mode inertia reduces their surface amplitude), these so-called mixed modes are of great diagnostic potential because they probe the stellar core and are very sensitive to stellar age.

\section{On extracting estimates of mode frequencies}\label{strategy}

\subsection{Overview of the different fitting strategies}
We computed the power density spectrum (PDS) of the time series based on the implementation of the Lomb-Scargle periodogram \citep{Lomb,Scargle} presented in \citet{PressRybicki}. This algorithm carries out reverse interpolation of the data onto a regular mesh and subsequently employs the fast Fourier transform. The power spectrum was then calibrated so that it satisfies Parseval's theorem, i.e., so that the total power in the positive-frequency side of the spectrum is equal to the variance of the time series (single-sided calibration). The effect of the window function is further taken into account when normalizing the PDS.

A total of eleven individual fitters (A2Z\_CR, A2Z\_DS, A2Z\_RG, AAU, IAS\_OB, IAS\_PG, IAS\_TA, OCT, ORK, QML and SYD) extracted estimates of the p-mode frequencies for at least one of the two stars and subsequently uploaded their results to the Cat Basket\footnote{\url{http://bison.ph.bham.ac.uk/kcatbasket/}} data exchange facility. Different fitting strategies have been adopted and sometimes the same fitting strategy has been applied in an independent manner. All the fitting strategies adopted are, however, based on Fourier methods, the main idea behind them being either the maximization of the likelihood of a multi-parameter model describing the data or a classic prewhitening method.

A frequency-domain representation of the data aims at modeling the limit PDS of the time series. Such a model typically includes a sum of symmetric Lorentzian profiles meant to describe the individual p modes, together with a flat term and a number of additional terms describing both instrumental and stellar background noise \citep{Anderson}: 
	\begin{equation}\label{freq_model}
		P(\nu)=\sum\limits_{n,l,m} \frac{H_{nl m}}{1+\left[2(\nu-\nu_{nl m})/\Gamma_{nl m}\right]^2}+B(\nu) \, ,
	\end{equation}
where $H$ is the mode height, $\Gamma$ is the mode linewidth (related to the mode lifetime or amplitude e-folding time, $\tau_{\rm{mode}}$, through $\pi\Gamma\!=\!1/\tau_{\rm{mode}}$), and $B(\nu)$ represents the background signal. The components arising from the decay of active regions, granulation and faculae are commonly represented using a Harvey-like model \citep{Harvey,Aigrain}:
	\begin{equation}
		B(\nu)=\sum\limits_{k}\frac{4\,\sigma_k^2\tau_k}{1+(2\pi\nu\,\tau_k)^{s_k}}+W \, ,
	\label{Harvey}	
	\end{equation}
where $\{\sigma_k\}$ are the amplitudes, $\{\tau_k\}$ are the characteristic timescales, $\{s_k\}$ are the slopes of the individual power laws in the denominator, and $W$ is a constant representing white noise (mainly due to photon shot noise). Equations \ref{freq_model} and \ref{Harvey} are representative of the models employed and individual fitters were further allowed to customize their own models. Hence we cannot talk of a reference model, as was done in \citet{Appourchaux08}. 

Most fitters opted for performing a global fit \citep[e.g.,][]{Appourchaux08} -- whereby the whole set of free parameters needed to describe the observed spectrum was optimized simultaneously -- while there was still room for pseudo-global (or local) fitting \citep[e.g.,][]{solarFLAG}, an approach traditionally adopted for Sun-as-a-star data, whereby narrow frequency windows are considered at a time. 

Statistical inference from the data was either frequentist or Bayesian in nature. The former approach relies on the straightforward application of a maximum likelihood estimator (MLE) taking into account the correct statistics of the spectrum \citep[e.g.,][]{MLE,Appourchaux08} or, in the special case of Gaussian statistics, a least squares estimator (LSE). To be specific: A2Z\_CR applied a global LSE to a smoothed version of the spectrum; A2Z\_DS applied a local MLE; IAS\_TA, OCT and QML applied a global MLE. The latter approach makes it possible to incorporate relevant prior information through Bayes' theorem. A regularized version of a MLE, known as maximum a posteriori \citep[MAP; e.g.,][]{Gaulme}, is a Bayesian point estimation method and was applied (globally) by A2Z\_RG and IAS\_PG. AAU and IAS\_OB also performed global fits but instead used Markov chain Monte Carlo (MCMC) techniques to map the posterior probability distributions of the frequency parameters \citep[e.g.,][]{Benomar,Gruberbauer,Hand&Camp}. With the exception of AAU and IAS\_OB, who estimated the frequency uncertainties based on the posterior distributions of these same parameters, all the remaining fitters provided formal error bars derived from the inverse Hessian matrix.         

ORK and SYD adopted a different approach, which did not involve fitting a model to the power density spectrum. ORK applied a procedure known as iterative sine-wave fitting \citep[ISWF; e.g.,][]{Bedding07,Alfio,White}: it iteratively removes sinusoidal components from the data, which are identified as the maxima of the Fourier spectrum of the residuals. Since the number of oscillation modes present in the data is unknown a priori, the algorithm requires a stopping rule that associates some degree of confidence to the amplitude of each extracted sinusoidal component. For the analysis by SYD, the power spectrum was smoothed by convolving with a Gaussian with full width at half maximum of $1.4\:{\rm{\mu Hz}}$ (comparable to the intrinsic linewidth of the modes) and the frequencies of the highest peaks were measured. Mode identification (assigning $l$ values) was done using the \'echelle diagram \citep{echelle} both by ORK and SYD. Finally, frequency uncertainties are estimated either analytically (ORK) or by performing Monte Carlo simulations with artificial data (SYD).

\subsection{Procedure for selecting frequency lists}\label{procedure}
A procedure for selecting relevant frequency lists to be used in stellar modeling has been described in \citet{Gemma}. However, we have opted for revising that procedure. This has been motivated, at first, by the fact that different fitting strategies provided frequency uncertainties that differed greatly. For instance, uncertainties that had been underestimated would undesirably upweight the respective individual frequency sets. Consequently, a more robust method of frequency selection was needed. Other refinements were also made and we give below a detailed description of the new, revised procedure.

We aim at selecting, for a given star and for a given mode degree identification, two frequency sets -- a \emph{minimal frequency set} and a \emph{maximal frequency set} -- that will provide initial constraints for the modeling and allow for refined model-fitting, respectively. As before, we would like to determine individual sets, as opposed to averaged sets, meaning that these sets are fully reproducible. 

We start by constructing both a \emph{minimal list} and a \emph{maximal list} of modes. Take $N$ as being the total number of individual fitters providing peak-bagging results for a given star and assume for now that $N\!>\!2$. For each $\{n,l\}$ pair, we apply Peirce's criterion \citep[see Appendix \ref{Peirceappend} for its implementation;][]{Peirce1,Peirce2} for the rejection of outliers and assess how many frequency estimates are retained. Inclusion or not of the mode in the \emph{minimal list} then results from a vote including all ($N$) fitters: If the number of frequencies retained is greater than\footnote{$\lfloor x \rfloor$ returns the closest integer less than or equal to $x$.} $\lfloor N/2 \rfloor$ then the mode is added to the \emph{minimal list}. Inclusion or not of the mode in the \emph{maximal list} results instead from a vote only including fitters providing estimates for that particular mode: If the number of frequencies retained is at least 2 then the mode is added to the \emph{maximal list}. The \emph{minimal list} is thus a subset of the \emph{maximal list}. For $N\!=\!2$ the \emph{minimal} and \emph{maximal lists} are degenerate and will coincide.

In the final stage of the procedure, we compute for each of the $N$ individual frequency sets the normalized root-mean-square deviation (nrmsd) with respect to the frequencies averaged over all contributing fitters, $\{\bar \nu_{nl}\}$, belonging to the \emph{minimal list} of modes:
	\begin{equation}
		\rm{nrmsd}=\sqrt{\frac{\sum\limits_{n,l} (\nu_{nl}-\bar \nu_{nl})^2/\sigma^2_{nl}}{N_1}} \, ,
	\end{equation}
where $\sigma_{nl}$ is the uncertainty in $\nu_{nl}$, and $N_1$ is the number of modes in a particular individual set that actually belong to the \emph{minimal list}. The \emph{best fit} is defined as being the individual set with the smallest normalized rms deviation. Note that by \emph{best fit} we mean here the most representative fit among the $N$ available sets and not necessarily the one closest to the truth. The \emph{minimal} and \emph{maximal frequency sets} are finally given by those modes provided by the \emph{best fit} that belong to the \emph{minimal} and \emph{maximal lists}, respectively. The homogeneous character of the revised procedure is reassured by generating the \emph{minimal} and \emph{maximal frequency sets} from the same individual set. The corresponding frequency uncertainties are simply those associated with the \emph{best fit}. When $N\!=\!2$ a \emph{single frequency set} is defined, the same happening for $N\!=\!1$ in which case it coincides with the only available individual set.

Although having been revised, this procedure is still subject to future improvement. The main reason for this is the fact that we assume the $l$ values to be correctly assigned to modes by all the fitters. A way of overcoming this would be to implement a clustering algorithm that groups frequencies and posteriorly tags them with a mode degree based on a vote. Another issue concerns the possibility that the \emph{best fit} misses some modes from the \emph{minimal} and \emph{maximal lists}. A solution for this would be to ask the fitter providing the \emph{best fit} to reanalyse the power spectrum. One last drawback of this procedure is the possibility that the \emph{minimal} and \emph{maximal frequency sets} may contain some outliers, i.e., frequency estimates that have been ruled out according to Peirce's criterion.

\section{Data analysis}\label{analysis}

\subsection{Time series preparation}\label{preparation}

The stars were observed from 2009 May 2 to 2010 March 19, i.e., from Quarter 0 (Q0) to Quarter 4 (Q4). The duty cycle over the course of the approximately 10 months of observations was above 90\%. Available time series suffered from several instrumental perturbations, and we have thus decided to develop our own corrections \citep{Rafacorrection}, independently from the pre-search data conditioning module of the \emph{Kepler} Science Pipeline \citep{Jenkins}, which generates corrected light curves for transiting-planet search.

\begin{figure}[t]
	\centering
	\includegraphics[width=0.36\textwidth,angle=90]{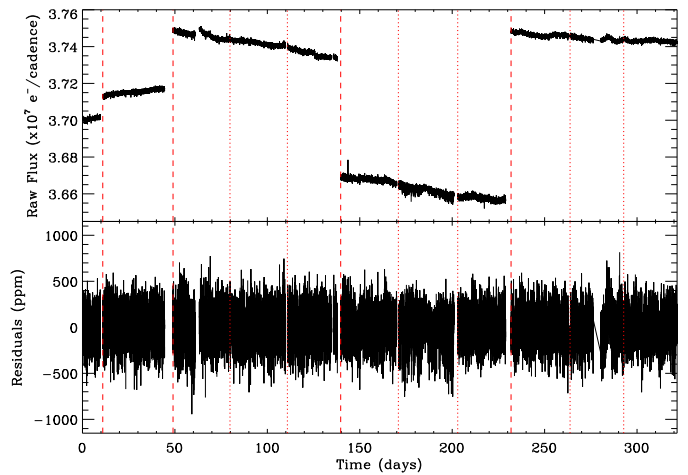}
	\caption{Raw-flux time series corrected only for outliers (top panel) and corrected relative time series (bottom panel) for the case of KIC~10273246. Vertical dashed lines mark the beginning of each Quarter, whereas vertical dotted lines separate the segments within a Quarter.}
	\label{fig_data}
\end{figure}

\begin{figure}[t]
	\centering
	\includegraphics[width=0.5\textwidth]{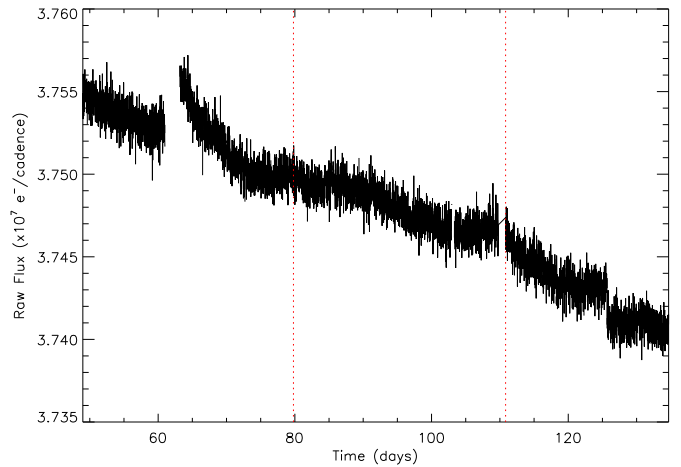}
	\caption{Raw-flux time series (corrected only for outliers) of Quarter 2 for the case of KIC~10273246. Vertical dotted lines separate the three segments of Q2. The induced thermal drift after a safe-mode event is visible between the 60th and 70th days.}
	\label{fig_q2}
\end{figure}

The raw-flux time series were corrected for three types of effects: outliers, jumps and drifts. Data points considered to be outliers exhibit point-to-point deviations greater than $3\sigma_{\rm{diff}}$, where $\sigma_{\rm{diff}}$ is the standard deviation of the first differences of the time series. This correction removed approximately 1\% of the data points. Jumps are sudden changes in the mean value of the time series caused by, e.g., attitude adjustments or a drop in pixel sensitivity (see Fig.~\ref{fig_data}). They were identified as causing spurious differences in the mean power of contiguous bins spanning one day. In cases where we know the photometric apertures to have changed, we by default check for jumps (only implemented in segments Q2.1 and Q4.1). Each jump has been manually validated. Finally, drifts are small, low-frequency perturbations due to temperature changes (e.g., after a long safe-mode event) and lasting for a few days (see Fig.~\ref{fig_q2}). The corrections are based on the software developed to deal with high-voltage perturbations in the GOLF/\emph{SoHO} instrument \citep{Rafa05}. We fit a 2nd- or 3rd-order polynomial function to the section of the time series where a thermal drift has been observed after comparing several light curves of the same Quarter. The fitted polynomial is then subtracted and another polynomial function of 1st- or 2nd-order -- used as a reference -- is added, which has been computed based on the observations done before and after the affected section. If the correction has to be applied to a border of the time series, then only one side of the light curve is processed. 

Once the aforementioned corrections have been applied, we merge the data of the different Quarters into a single time series, after equalizing the average counting-rate level between the Quarters (or sometimes even within a Quarter when some instrumental parameters have been changed). In order to do so, and to allow conversion into units of ppm, we use a series of 6th-order polynomial fits, one for each segment. Finally, we normalize the standard deviation of the data obtained during Q3 to the average of the other Quarters, since Q3 is considerably noisier for the two stars. This normalization proved to be a good compromise between using the noisy Q3 data and not using them at all. Given the insufficient technical information available, it is not possible to know whether or not there is a variation in the gain of the CCD module during Q3. If such a gain variation is indeed present, the mode amplitudes would be affected and the adopted normalization will then correct these amplitudes toward something closer to their real value. If, on the other hand, no gain variation is present, then the adopted correction will effectively reduce the mode amplitudes. Having reduced the standard deviation of the data in Q3 by 34\% for KIC~10273246 and 49\% for KIC~10920273, and given that Q3 represents 3/7 of the total length of the time series used for peak-bagging (i.e., from Q0 to Q3; see Sect.~\ref{modef}), then mode amplitudes would be reduced by about 14\% and 21\%, respectively. However, we believe this latter scenario to be less likely and hence the quoted values of 14\% and 21\% can only be regarded as upper limits to mode amplitude reduction, which in reality may be much smaller.

\subsection{Rotational modulation}\label{rotmod}
\begin{figure}[t]
	\centering
	\includegraphics[width=0.5\textwidth]{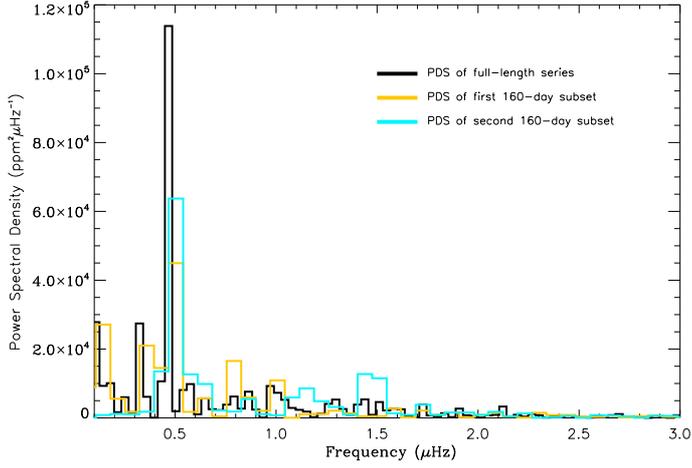}
	\caption{Low-frequency end -- between 0.1 and $3\:{\rm{\mu Hz}}$ -- of the power density spectrum of KIC~10273246.}
	\label{rot_muld}
\end{figure}

\begin{figure}[t]
	\centering
	\includegraphics[width=0.5\textwidth]{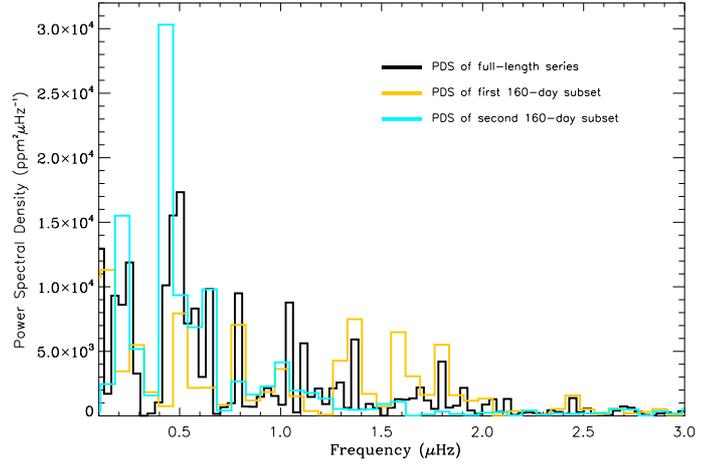}
	\caption{Similar to Fig.~\ref{rot_muld} but for the case of KIC~10920273.}
	\label{rot_scu}
\end{figure}

To investigate the stellar rotational period ($P_{\rm{rot}}$) of these stars we seek high-SNR peaks in the low-frequency end of the PDS. However, the procedure for merging the different data sets described in the previous paragraph filters out the power density below $1\:{\rm{\mu Hz}}$. Therefore, starting with the corrected data, we have generated new merged time series in which a triangular smoothing -- over a period selected from the range of 12--20 days -- has been used to normalize the light curves instead of the 6th-order polynomial fit \citep[for details, see][]{Rafacorrection}. Finally, we computed the PDS of the full-length time series and also of two subsets of 160 days each. The low-frequency ends of the PDS are shown in Figs.~\ref{rot_muld} and \ref{rot_scu} for KIC~10273246 and KIC~10920273, respectively. A different pattern of peaks appears for each star. In the case of KIC~10273246 (Fig.~\ref{rot_muld}), the highest peak is seen at about $0.50\:{\rm{\mu Hz}}$ ($P_{\rm{rot}}\!\approx\!23\:{\rm{days}}$) in the three spectra considered, without any signature of differential rotation. This turns out to be a reliable signature of rotational modulation, since comparison with the power spectra of other stars observed with the same CCD module ruled out the possibility of an instrumental artifact. On the other hand, KIC~10920273 (Fig.~\ref{rot_scu}) exhibits a high peak at about $0.43\:{\rm{\mu Hz}}$, though mostly during the second subset of 160 days. Indeed, this peak is not very significant, and during the first subset it is at the same level as that of the adjacent peaks. The PDS of the full-length time series also reveals itself as being rather noisy, and it is difficult to disentangle the stellar rotational period from it. Therefore, while a cyclic rotational frequency of $0.43\:{\rm{\mu Hz}}$ ($P_{\rm{rot}}\!\approx\!27\:{\rm{days}}$) seems plausible, we cannot exclude the possibility that this peak is associated with the roll schedule of the telescope. One of several alternative explanations for the low-SNR peaks would be that this star was observed during a low magnetic-activity period, with a small number of spots on its surface. A deeper study based on a longer data set is necessary to confirm the rotational rate of KIC~10920273.

\subsection{Power spectral density of the background}
\begin{figure*}
   \centering
   \includegraphics{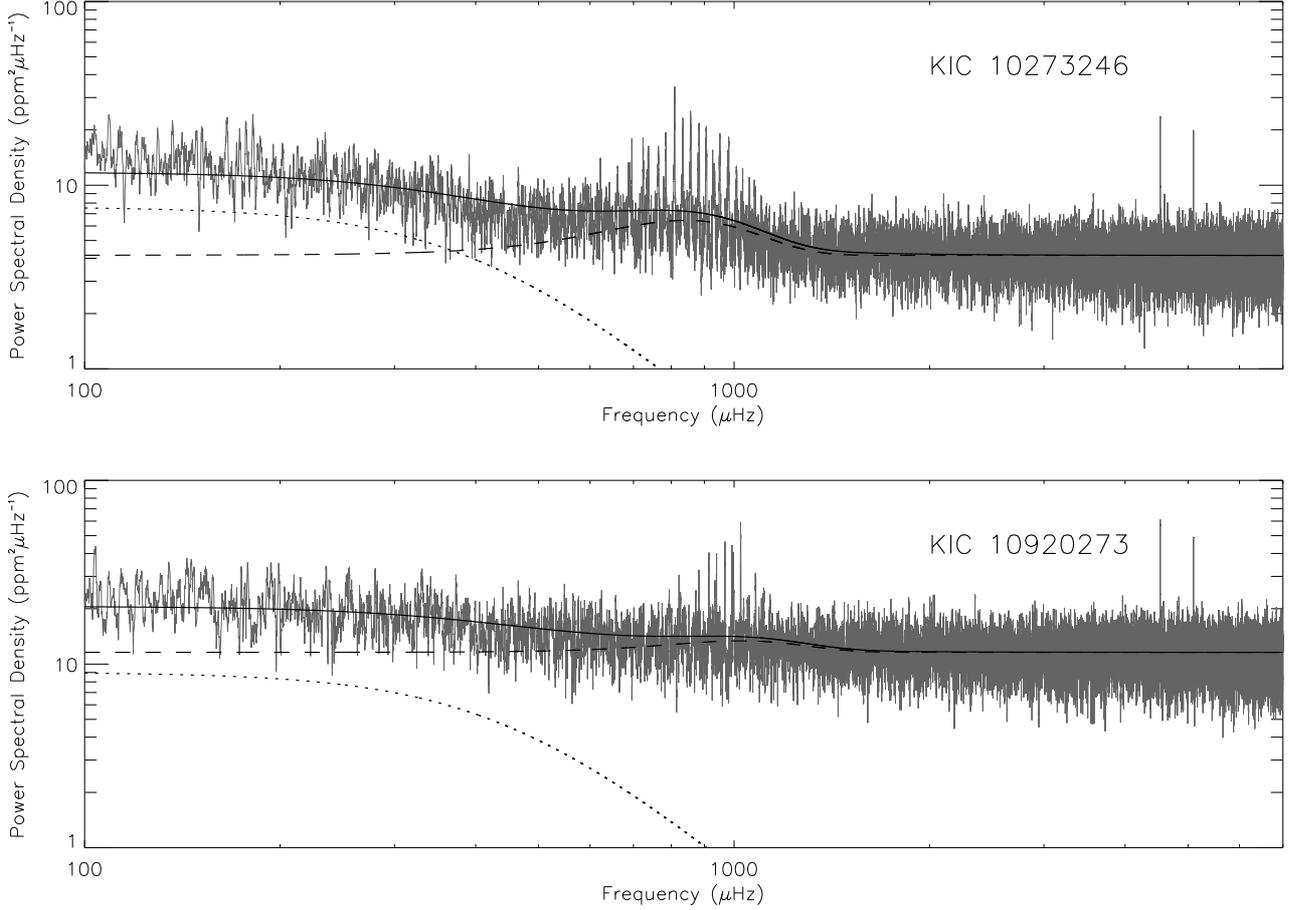}
   \caption{Power density spectra -- smoothed over $1\:{\rm{\mu Hz}}$ in order to enhance p-mode visibility -- of KIC~10273246 (top panel) and KIC~10920273 (bottom panel), plotted on a log-log scale. The solid black lines represent the fits described in the text. The remaining lines represent different components of the fitting model: the power envelope due to oscillations added to the offset from white noise (dashed), and granulation (dotted).}
   \label{PDS_bckg}
\end{figure*}

The power density spectra -- based on the first seven months of observations, i.e., from Q0 to Q3 -- of KIC~10273246 and KIC~10920273 are displayed in Fig.~\ref{PDS_bckg}. The background signal rises toward the low-frequency end of the spectra with contributions from granulation and activity. At the high-frequency end, the spectra are dominated by photon shot noise. Also visible at the high-frequency end are a number of peaks corresponding to harmonics of the inverse of the long-cadence period, an artifact appearing in power spectra of short-cadence time series \citep{Gilliland} but having negligible influence on the current study. In between these two frequency regimes there is a conspicuous cluster of power due to the presence of p modes, with the hotter and brighter target KIC~10273246 displaying the higher SNR.

We fitted a model similar to the one described in Eq.~\ref{Harvey} to a heavily smoothed version of both spectra. This model included an additional Gaussian function aimed at describing the p-mode power-excess hump \citep[e.g.,][]{Kallinger}. The fitting window was from $100\:{\rm{\mu Hz}}$ up to the Nyquist frequency, hence we did not consider a term accounting for the decay of active regions, whose typical timescale is considerably longer. We found no evidence for a facular component based on a simple visual inspection of the spectra, thus having not included such a component in the background model. Faculae had previously been reported on a couple of \emph{Kepler} solar-type targets by \citet{Kp_Chaplin}. A component carrying the signature of stellar granulation is, however, clearly displayed by both stars. Figure \ref{PDS_bckg} displays the fitted models, while values of the fitted parameters are given in Table \ref{Parameters}. Note that the slope of both granulation components has been fixed at the same value, i.e., $s_{\rm{gran}}\!=\!3$, which is closer to solar \citep[e.g.,][]{Michelslope} than the value of 2 originally proposed by \citet{Harvey}. Although we lack a physical reason for fixing this slope, the intention is to facilitate the comparison between the values of $\tau_{\rm{gran}}$.
 
The high-frequency noise power spectral density ($W$) is approximately 14\% and 16\% higher than predictions in the cases of KIC~10273246 and KIC~10920273, respectively. These predictions have been computed according to the empirical minimal term model for the noise presented in \citet{Gilliland}, which takes into account the \emph{Kepler}-band magnitude of the star and the performance of the instrument. Noise levels are thus close to being Poisson-limited. A likely source of this extra noise is the larger scatter in Q3. 

\begin{table}[h]
\caption{\label{Parameters}Background model parameters describing non-resonant features.}
\centering
\begin{tabular}{lcccc}
\hline\hline
Star&$\sigma_{\rm{gran}}$&$\tau_{\rm{gran}}$&$s_{\rm{gran}}$\tablefootmark{a}&$W$\\
&$(\rm{ppm})$&(s)&&$(\rm{ppm^2\,\mu Hz^{-1}})$\\
\hline
KIC~10273246&$69.9\pm0.8$&$390\pm12$&3&$4.15\pm0.02$\\
KIC~10920273&$80.1\pm0.4$&$351\pm4$&3&$11.58\pm0.02$\\
\hline
\end{tabular}
\tablefoot{\\
\tablefoottext{a}{The slope of both Harvey-like terms has been fixed.}
}
\end{table}

\subsection{The global asteroseismic parameters $\Delta\nu$, $\nu_{\rm{max}}$ and $\delta\nu_{n0}$}
In the past few years, a number of automated pipelines have been developed to measure global asteroseismic parameters of solar-like oscillators \citep{Huber,MA09,Roxburgh,Campante,Hekker,Karoff10,Mathur}. Most of them have already been successfully tested on \emph{CoRoT} data \citep[e.g.,][]{HD170987}. The automated nature of these pipelines is required if we are to efficiently exploit the plenitude of data made available by \emph{Kepler} on these targets. A thorough comparison of complementary analysis methods used to extract global asteroseismic parameters of main-sequence and subgiant solar-like oscillators is presented in \citet{Verner}.

\begin{figure}[t]
   \centering
   \includegraphics[width=0.5\textwidth]{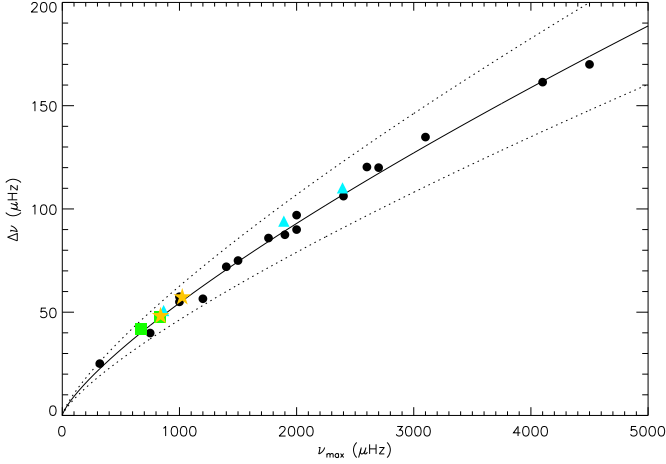}
   \caption{Observed relation between $\Delta\nu$ and $\nu_{\rm{max}}$. Here we reproduce fig.~2 of \citet{Stello2} in linear scale, where only stars whose $\nu_{\rm{max}}$ is greater than the Nyquist frequency for long-cadence sampling are displayed. The solid line is the power-law fit in their eq.~1 with the corresponding $\pm15\%$ deviations shown as dotted lines. We added to the plot the two stars being considered in this study (five-pointed stars), the two stars (squares) of \citet{FurryMathur}, and the three bright G-type stars (triangles) of \citet{Kp_Chaplin}.}
   \label{Dnu_numax}
\end{figure}

The results of the different pipelines on global parameter extraction have also been uploaded to the Cat Basket. For a given global parameter, some groups submitted results using more than one method or on more than one data release. In Table \ref{global} we give a representative set of estimates of $\Delta\nu$ and $\nu_{\rm{max}}$ for the two stars in our study. These estimates were supplied by one of the pipelines, namely, the one described in \citet{MA09}, and are based on the analysis of the full-length time series. The large separation is a function of frequency. Consequently, the extent of variation of its mean value will depend upon the variation in frequency range adopted for its computation. However, as long as this frequency range includes $\nu_{\rm{max}}$, the impact of small differences in the range can be neglected. For the sake of completeness, we give the adopted frequency ranges in Table \ref{global}. An observed relation between $\Delta\nu$ and $\nu_{\rm{max}}$ for solar-like oscillations in main-sequence stars is presented in \citet{Stello2}:
	\begin{equation}
		\Delta\nu \propto \nu_{\rm{max}}^{0.77} \, .
	\end{equation}
The values quoted in Table \ref{global} for these two global parameters satisfy this relation, as can be seen in Fig.~\ref{Dnu_numax}. Also, $\tau_{\rm{gran}}$ is seen to scale inversely with $\nu_{\rm{max}}$, as predicted by \citet{KB11}.

The same pipeline has provided estimates of $\delta\nu_{n0}$ based on the analysis of seven months of data. These are also given in Table \ref{global}.

\begin{table}[h]
\caption{\label{global}Estimates of the global asteroseismic parameters $\Delta\nu$, $\nu_{\rm{max}}$ and $\delta\nu_{n0}$.}
\centering
\begin{tabular}{lcccc}
\hline\hline
Star&$\langle\Delta\nu\rangle$&Range&$\nu_{\rm{max}}$&$\langle\delta\nu_{n0}\rangle$\\
&$(\rm{\mu Hz})$&$(\rm{\mu Hz})$&$(\rm{\mu Hz})$&$(\rm{\mu Hz})$\\
\hline
KIC~10273246&$48.2\pm0.5$&$[537,1140]$&$839\pm51$&$5.6\pm1.2$\\
KIC~10920273&$57.3\pm0.8$&$[757,1290]$&$1024\pm64$&$6.0\pm1.5$\\
\hline
\end{tabular}
\end{table}

\subsection{Mode frequencies}\label{modef}
Estimates of mode frequencies were obtained by several individual fitters for both stars, based on the first seven months of observations (from Q0 to Q3). The procedure described in Sect.~\ref{procedure} was then used to select \emph{minimal} and \emph{maximal frequency sets}. An ad hoc step was, however, implemented at this stage, that aimed at removing doubtful modes from these sets based on SNR considerations. This quality control led to the removal of a few modes at the low- and high-frequency ends of the sets.

Another issue relates to the presence of mixed modes in the power spectrum. As we will see below, the \emph{best fit} relies, for both stars, on an algorithm based on a frequency-domain representation of the data (cf.~Eq.~\ref{freq_model}). Moreover, the deterministic models that have been used only contained modes of degree up to $l\!=\!2$, with three modes per radial order, one of each degree. The presence of an extra mixed mode in the vicinity of a mode with low SNR therefore greatly complicates the fit due to the inadequacy of the model. As a result, we opted for prewhitening the longer-lived mixed modes and give their frequencies a posteriori, while discarding any neighboring power structure with low SNR fitted by the MLE-based algorithm.

\begin{figure}[t]
   \centering
   \includegraphics[width=0.5\textwidth]{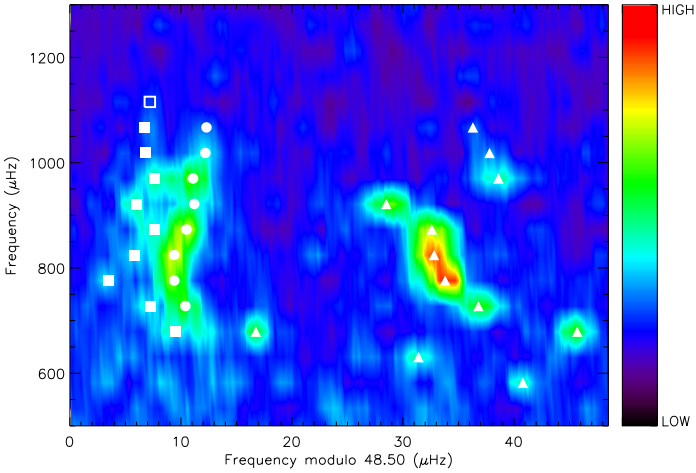}
   \caption{\'Echelle diagram of the power density spectrum of KIC~10273246 (the colorbar decodes the power density level). The \emph{minimal} (filled symbols) and \emph{maximal} (filled and open symbols) \emph{frequency sets} are displayed. Symbol shapes indicate mode degree: $l\!=\!0$ (circles), $l\!=\!1$ (triangles) and $l\!=\!2$ (squares).}
   \label{Mulder_echelle}
\end{figure}

Eleven individual fitters (A2Z\_CR, A2Z\_DS, A2Z\_RG, AAU, IAS\_OB, IAS\_PG, IAS\_TA, OCT, ORK, QML and SYD) 
have provided results for KIC~10273246. A summary of the anonymous \emph{best fit} is given in Table \ref{extrainfo}. The fitter has also imposed priors on the parameters used to describe the background signal, which were based on the output of a previous fit to the background alone. Table \ref{Freq_Mulder} and Fig.~\ref{Mulder_echelle} combine the \emph{minimal} and \emph{maximal frequency sets} (recall that the former is a subset of the latter). We have identified a total of 30 modes of degree $l\!=\!0,1,2$ that span a frequency range of width $\approx\!10\,\Delta\nu$. The mode at $1122.70\:{\rm{\mu Hz}}$ has only been reported by two fitters, who have identified it as being a radial mode despite its alignment with the $l\!=\!2$ ridge. Given its very low power and the known convergence issues experienced by MLE-based methods under such low SNR conditions, we have decided to change its tagging a posteriori from $l\!=\!0$ to $l\!=\!2$. This mode should thus be considered with caution.

The problem of ridge identification (i.e., the tagging of modes by degree $l$) in F stars dates back to when \emph{CoRoT} observations of HD~49933 were first analysed by \citet{Appourchaux08}. In the present case such an identification can be simply done by visual inspection of the \'echelle diagram in Fig.~\ref{Mulder_echelle}. Nonetheless, two of the fitters (IAS\_OB and IAS\_TA) provided results for the two complementary identification scenarios. From the results provided by IAS\_TA we are able to compute the likelihood ratio \citep[e.g.,][]{Appourchaux98} in favor of our preferred scenario, whereas computation of its Bayesian counterpart, the more conservative Bayes' factor \citep[e.g.,][]{Liddle,Hand&Camp}, is possible based on the results provided by IAS\_OB. Both approaches returned conclusive values in support of the adopted scenario.  

\begin{figure}[t]
   \centering
   \includegraphics[width=0.5\textwidth]{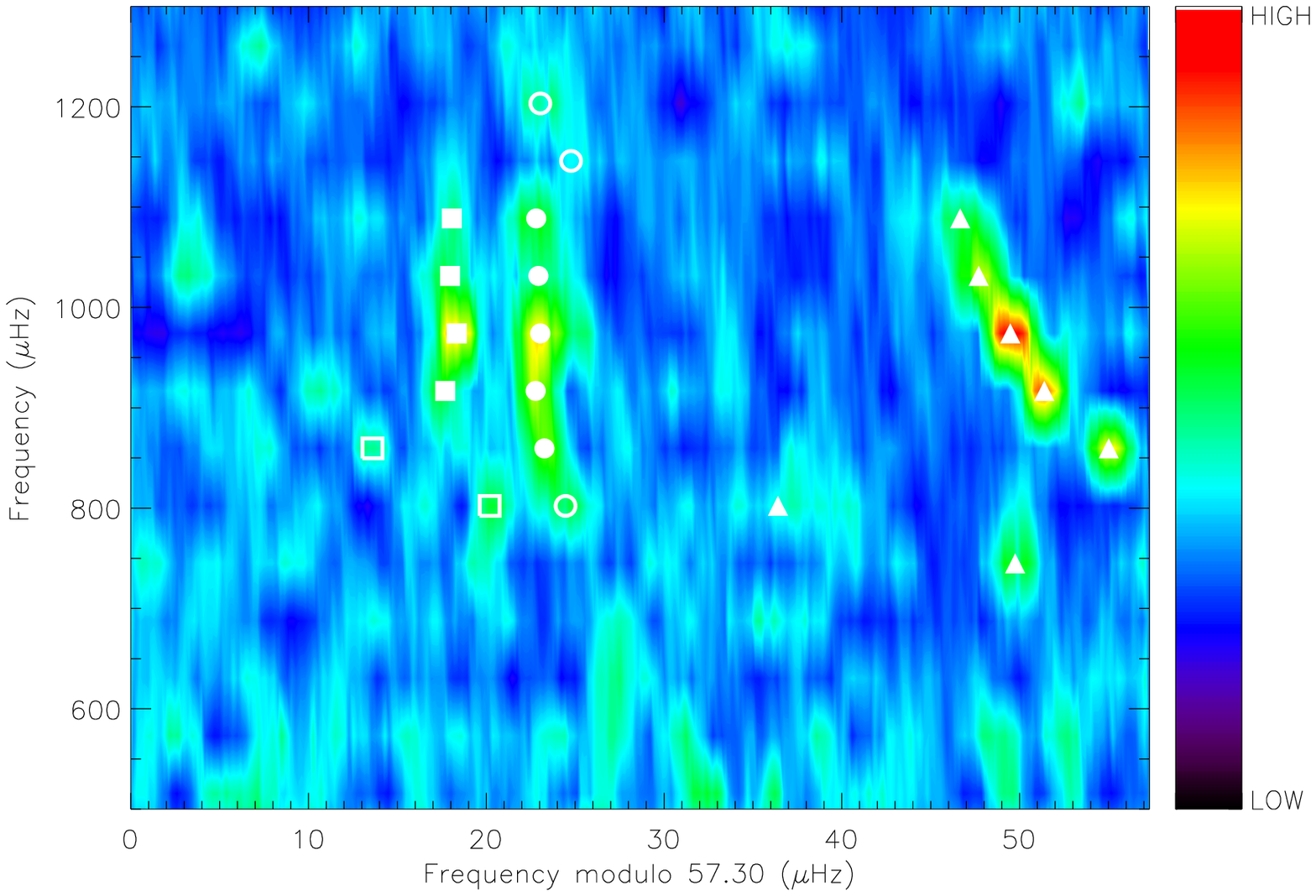}
   \caption{Similar to Fig.~\ref{Mulder_echelle} but for the case of KIC~10920273.}
   \label{Scully_echelle}
\end{figure}

Ten individual fitters (A2Z\_CR, A2Z\_DS, A2Z\_RG, IAS\_OB, IAS\_PG, IAS\_TA, OCT, ORK, QML and SYD) have provided results for KIC~10920273. A summary of the anonymous \emph{best fit} is given in Table \ref{extrainfo}. In addition to the information provided in that table, it should also be mentioned that a Gaussian prior has been imposed on $\delta\nu_{n0}$. Table \ref{Freq_Scully} and Fig.~\ref{Scully_echelle} combine the \emph{minimal} and \emph{maximal frequency sets}. We have identified a total of 21 modes -- considerably less than for KIC~10273246 -- of degree $l\!=\!0,1,2$ that span a frequency range of width $\approx\!8\,\Delta\nu$. The $l\!=\!1$ mode at $1135.36\:{\rm{\mu Hz}}$ is very close to the second harmonic ($2 \times 566.391\:{\rm{\mu Hz}}$) of the long-cadence sampling. Although this anomaly is known to be more prominent for intermediate harmonics, we nonetheless follow the recommendation of \citet{Gilliland} to flag this mode as suspect.

\begin{table*}[!t]
\caption{\label{extrainfo}Summary of the \emph{best fits}.}
\centering
\begin{tabular}{lccccc}
\hline\hline
Star&Method&Height ($H$)&Linewidth ($\Gamma$)&Splitting ($\nu_{\rm{rot}}$)&Inclination\tablefootmark{a} ($i$)\\
\hline
KIC~10273246&Global&One free parameter ($H_{l=0}$) per order&One free parameter per order&Free&Free\\
&MAP&No prior&No prior&No prior&No prior\\
&&$H_{l=1}/H_{l=0}\!=\!1.5$ (fixed)&&&\\
&&$H_{l=2}/H_{l=0}\!=\!0.5$ (fixed)&&&\\ \\
KIC~10920273&Global&One free parameter ($H_{l=0}$) per order&One free parameter per order&Fixed at $0\:{\rm{\mu Hz}}$&Fixed at $0^\circ$\\
&MAP&No prior&Gaussian prior&&\\
&&$H_{l=1}/H_{l=0}\!=\!1.5$ (fixed)&&&\\
&&$H_{l=2}/H_{l=0}\!=\!0.5$ (fixed)&&&\\
\hline
\end{tabular}
\tablefoot{\\
\tablefoottext{a}{Introduced in Eq.~\ref{GS}.}
}
\end{table*}

The quasi-regularity of the small ($\delta\nu_{n0}$) and large frequency separations ($\Delta\nu_{n0}$ and $\Delta\nu_{n2}$) is evident from Figs.~\ref{Mulder_echelle} and \ref{Scully_echelle}. Notice that if these stars were to strictly obey the asymptotic relation in Eq.~\ref{asymptotic}, they would then exhibit vertical ridges in the \'echelle diagram provided use of the correct $\Delta\nu$. The small separation $\delta\nu_{n0}$ is, however, more clearly distinguished in the case of KIC~10920273, which might be an indication of smaller mode linewidths in this cooler star (see Sect.~\ref{WH} for a discussion on mode linewidths).

\begin{figure}[t]
   \centering
   \includegraphics[width=0.5\textwidth]{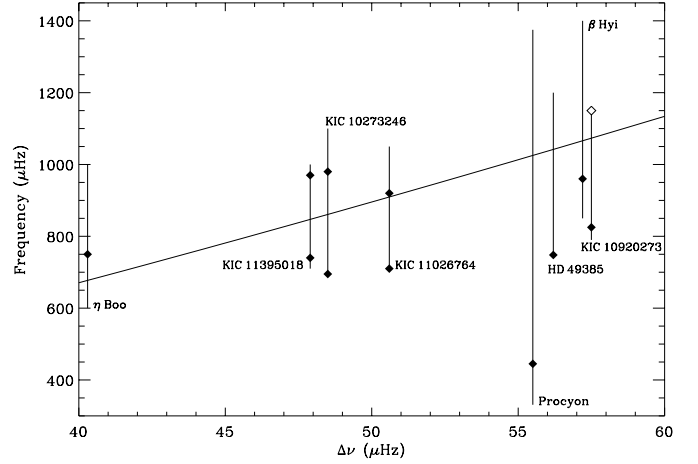}
   \caption{p-g diagram showing the frequencies of avoided crossings (diamonds) plotted against the large separation for a number of stars. The vertical lines show the range over which the $l\!=\!1$ ridge is clearly seen. The inclined line is the observed relation between $\Delta\nu$ and $\nu_{\rm{max}}$ of \citet{Stello2}. For KIC~10273246 there are two clear avoided crossings, whereas for KIC~10920273 there is only one clear crossing plus another likely one (open symbol).}
   \label{pg}
\end{figure}

A striking feature in both \'echelle diagrams is the jagged appearance of the $l\!=\!1$ ridge, a trademark of the presence of avoided crossings and an indicator of the evolved nature of these stars. These same features have also been seen in the cases of ground-based observations of $\eta$ Boo \citep{Kjeldsen03}, $\beta$ Hyi \citep{Bedding07} and possibly Procyon \citep{Procyon}, as well as in the cases of the \emph{CoRoT} target HD~49385 \citep{HD49385}, and KASC survey targets KIC~11026764 \citep{Gemma}, KIC~11395018 and KIC~11234888 \citep{FurryMathur}. Figure \ref{pg} displays a so-called p-g diagram as introduced by \citet{Bedding_pg}, where the frequencies of the avoided crossings (i.e., the frequencies of the pure g modes in the core cavity) for a number of stars are plotted against the large separation of the p modes. Much of the diagnostic potential of mixed modes can be captured in this way, since their overall pattern is determined by the mode bumping at each avoided crossing, which in turn is determined by the g modes trapped in the core. This diagram could prove to be an instructive way to display results of many stars and to allow for a first comparison with theoretical models. We also report here the possible presence of a $l\!=\!2$ mixed mode in the power spectrum of KIC~10920273 (at $873.10\:{\rm{\mu Hz}}$) that should, however, be confirmed by stellar models.

Detection of $l\!=\!3$ modes with photometric observations is made very difficult due to geometric cancellation effects. Solar-like oscillations with $l\!=\!3$ from \emph{Kepler} photometry have nonetheless been reported for a set of low-luminosity red giants by \citet{Bedding_rg}. \citet{HD49385} have also reported the presence of $l\!=\!3$ modes for the \emph{CoRoT} target HD~49385. We should bear in mind that, except for ORK and SYD, all the remaining fitters used deterministic models in their frequency-domain representations of the data that only contained modes of degree up to $l\!=\!2$, meaning that a statistical assessment of the presence or not of $l\!=\!3$ modes could not be done. ORK and SYD, which were the only fitters that did not make any prior assumptions about the degree of the modes, have not reported the detection of modes that could be interpreted as $l\!=\!3$ modes.

\begin{table}[t]
\caption{\label{Freq_Mulder}The \emph{minimal} and \emph{maximal frequency sets} of observed oscillation frequencies for KIC~10273246.}
\centering
\begin{tabular}{ccc}
\hline\hline
$l$&Frequency&Uncertainty\\
&$(\rm{\mu Hz})$&$(\rm{\mu Hz})$\\
\hline
0&737.90&0.30\\
0&785.40&0.20\\
0&833.90&0.20\\
0&883.50&0.20\\
0&932.70&0.50\\
0&981.10&0.30\\
0&1030.70&0.40\\
0&1079.30&0.20\\
\hline
1&622.80&0.20\\
1&661.90&0.50\\
1&695.75\tablefootmark{b}&0.27\\    
1&724.70&0.20\\
1&764.30&0.30\\
1&809.80&0.20\\
1&857.30&0.20\\
1&905.60&0.30\\
1&950.00&0.30\\
1&1008.60&0.40\\
1&1056.30&0.20\\
1&1103.30&0.40\\
\hline
2&688.50&0.70\\
2&734.80&0.60\\
2&779.50&0.40\\
2&830.30&0.40\\
2&880.60&0.50\\
2&927.50&0.40\\
2&977.60&0.40\\
2&1025.30&1.30\\
2&1073.70&0.20\\
2&1122.70\tablefootmark{a,c}&0.40\\
\hline
\end{tabular}
\tablefoot{\\
\tablefoottext{a}{Mode belonging exclusively to the \emph{maximal frequency set}.}\\
\tablefoottext{b}{$l\!=\!1$ mixed mode introduced a posteriori.}\\
\tablefoottext{c}{Tagging changed a posteriori from $l\!=\!0$ to $l\!=\!2$.}
}
\end{table}

\begin{table}[t]
\caption{\label{Freq_Scully}The \emph{minimal} and \emph{maximal frequency sets} of observed oscillation frequencies for KIC~10920273.}
\centering
\begin{tabular}{ccc}
\hline\hline
$l$&Frequency&Uncertainty\\
&$(\rm{\mu Hz})$&$(\rm{\mu Hz})$\\
\hline
0&826.66\tablefootmark{a}&0.25\\
0&882.77&0.20\\
0&939.58&0.16\\
0&997.14&0.18\\
0&1054.33&0.30\\
0&1111.51&0.25\\
0&1170.77\tablefootmark{a}&0.33\\
0&1226.34\tablefootmark{a}&0.33\\
\hline
1&794.65\tablefootmark{b}&0.32\\    
1&838.61\tablefootmark{b}&0.25\\ 
1&914.52&0.16\\
1&968.19&0.13\\
1&1023.58&0.14\\
1&1079.10&0.31\\
1&1135.36\tablefootmark{d}&0.31\\
\hline
2&822.39\tablefootmark{a}&0.28\\
2&873.10\tablefootmark{a,c}&0.32\\  
2&934.49&0.22\\
2&992.44&0.13\\
2&1049.36&0.39\\
2&1106.76&0.34\\
\hline
\end{tabular}
\tablefoot{\\
\tablefoottext{a}{Mode belonging exclusively to the \emph{maximal frequency set}.}\\
\tablefoottext{b}{$l\!=\!1$ mixed mode introduced a posteriori.}\\
\tablefoottext{c}{Possible $l\!=\!2$ mixed mode introduced a posteriori.}\\
\tablefoottext{d}{Mode very close to the second harmonic of the inverse of the long-cadence period.}
}
\end{table}


\subsection{A word on mode linewidths, heights, and amplitudes}\label{WH}
A thorough discussion of the mode linewidths, heights, and amplitudes goes beyond the scope of this work. However, there are some aspects we would like to mention here.

\begin{figure}[t]
   \centering
   \includegraphics[width=0.5\textwidth]{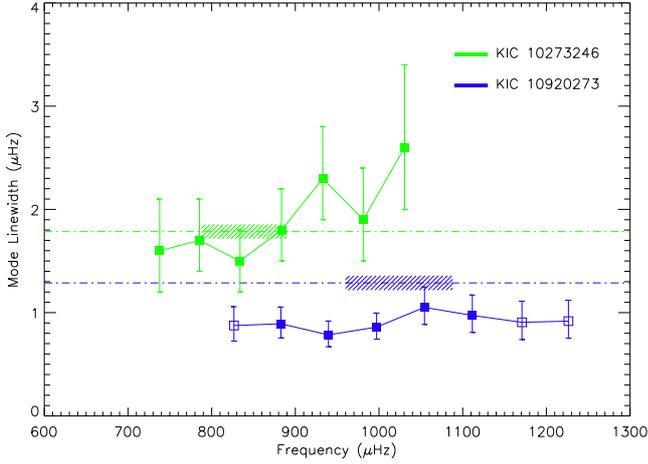}
   \caption{Linewidths of the radial modes returned for KIC~10273246 and KIC~10920273 by the respective \emph{best fit}. Modes represented by open symbols belong exclusively to the \emph{maximal frequency set}. Dot-dashed lines mark the predicted mean linewidths of the most prominent modes using Eq.~\ref{width_temp}. The horizontal dimension of the line-filled areas represents the uncertainty in $\nu_{\rm{max}}$.}
   \label{widths}
\end{figure}

The intrinsic frequency resolution of the spectra ($\approx\!0.05\:{\rm{\mu Hz}}$) makes it possible to resolve the modes. This condition is obeyed provided the observation length $T\!\gg\!2\tau_{\rm{mode}}$ \citep{Chaplin03}. Figure \ref{widths} displays, for each star, the linewidths of the radial modes returned by the respective \emph{best fit} (see also Tables \ref{Freq_Mulder2} and \ref{Freq_Scully2}). The radial modes considered are those belonging to the \emph{maximal frequency set}. Notice the near-constancy with frequency of the mode linewidths in the case of KIC~10920273, whereas for KIC~10273246 the linewidths increase steadily, although with considerably larger error bars (note that both sets of error bars were derived from the inverse Hessian matrix). Using pulsation computations of a grid of stellar models and the first asteroseismic results on mode lifetimes of solar-like stars, \citet{Chaplin09} suggested a simple scaling relation between the mean mode linewidth of the most prominent p modes and $T_{\rm{eff}}$:
	\begin{equation}\label{width_temp}
		\langle\Gamma\rangle \propto T_{\rm{eff}}^4 \, .
	\end{equation}
Figure \ref{widths} also displays the resulting predictions of mean linewidths of the most prominent modes (we have considered $\langle\Gamma\rangle_{\sun}\!\approx\!1.2\:{\rm{\mu Hz}}$ and $T_{\rm{eff}\,\sun}\!=\!5777\:{\rm{K}}$, and taken the recalibrated KIC temperatures). The agreement with the observed values is fairly good in the case of KIC~10273246. On the other hand, the predicted value obtained for KIC~10920273 using Eq.~\ref{width_temp} overestimates the observed linewidths. Assuming validity of this equation, this might be the result of the combination of two factors, namely, an overestimation of $T_{\rm{eff}}$ and an overestimation of the fitted background that leads to underestimated fitted linewidths. Nonetheless, the F-type star KIC~10273246 exhibits the larger mode linewidths, as expected from Eq.~\ref{width_temp}. We should also note that \citet{Baudin} found a much stronger dependence of the mean mode linewidth on $T_{\rm{eff}}$.

\begin{figure}[t]
   \centering
   \includegraphics[width=0.5\textwidth]{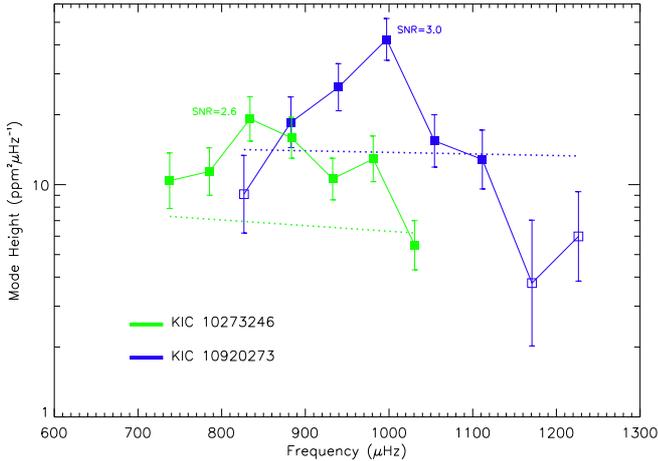}
   \caption{Heights of the radial modes returned for KIC~10273246 and KIC~10920273 by the respective \emph{best fit}, plotted on a log-linear scale. Modes represented by open symbols belong exclusively to the \emph{maximal frequency set}. Also indicated is the SNR of the strongest modes as well as the background fits of Fig.~\ref{PDS_bckg} (dotted lines).}
   \label{heights}
\end{figure}

When a mode is resolved, as is the case here, it is the mode height, $H$, that determines the SNR in power, viz., the height-to-background ratio. Figure \ref{heights} displays, for each star, the heights of the radial modes returned by the respective \emph{best fit} (see also Tables \ref{Freq_Mulder2} and \ref{Freq_Scully2}). The modes are the same as shown in Fig.~\ref{widths}. We also indicate in Fig.~\ref{heights} the SNR of the strongest radial modes, as well as the background fits of Fig.~\ref{PDS_bckg} (which are an indicator of ${\rm{SNR}}\!=\!1$). Overall, KIC~10273246 exhibits a higher SNR if we take into account the whole plotted frequency bands. Also apparent is the larger width of the p-mode hump of KIC~10920273, which roughly scales with $\nu_{\rm{max}}$ \citep[e.g.,][]{Stello07,MosserRG}.  


\begin{table*}[!h]
\caption{\label{Freq_Mulder2}Linewidths, heights, and bolometric amplitudes of radial modes for KIC~10273246 returned by the \emph{best fit} (these are not provided for the radial mode of highest frequency since convergence was not properly achieved).}
\centering
\begin{tabular}{ccccccc}
\hline\hline
Frequency&Linewidth&Uncertainty&Height&Uncertainty&Amplitude&Uncertainty\\
$(\rm{\mu Hz})$&$(\rm{\mu Hz})$&$(\rm{\mu Hz})$&$(\rm{ppm^2\,\mu Hz^{-1}})$&$(\rm{ppm^2\,\mu Hz^{-1}})$&$(\rm{ppm})$&$(\rm{ppm})$\\
\hline
737.90&1.6&$+0.5$/$-0.4$&10.4&$+3.3$/$-2.5$&5.48&$\pm0.81$\\
785.40&1.7&$+0.4$/$-0.3$&11.4&$+3.0$/$-2.4$&5.90&$\pm0.80$\\
833.90&1.5&$+0.3$/$-0.3$&19.2&$+4.7$/$-3.8$&7.17&$\pm0.75$\\
883.50&1.8&$+0.4$/$-0.3$&15.9&$+3.6$/$-2.9$&7.14&$\pm0.75$\\
932.70&2.3&$+0.5$/$-0.4$&10.6&$+2.4$/$-2.0$&6.60&$\pm0.77$\\
981.10&1.9&$+0.5$/$-0.4$&12.9&$+3.3$/$-2.6$&6.61&$\pm0.73$\\
1030.70&2.6&$+0.8$/$-0.6$&5.5&$+1.5$/$-1.2$&5.09&$\pm0.83$\\
1079.30&$\cdots$&$\cdots$&$\cdots$&$\cdots$&$\cdots$&$\cdots$\\
\hline
\end{tabular}
\end{table*}

\begin{table*}[!h]
\caption{\label{Freq_Scully2}Linewidths, heights, and bolometric amplitudes of radial modes for KIC~10920273 returned by the \emph{best fit}.}
\centering
\begin{tabular}{ccccccc}
\hline\hline
Frequency&Linewidth&Uncertainty&Height&Uncertainty&Amplitude&Uncertainty\\
$(\rm{\mu Hz})$&$(\rm{\mu Hz})$&$(\rm{\mu Hz})$&$(\rm{ppm^2\,\mu Hz^{-1}})$&$(\rm{ppm^2\,\mu Hz^{-1}})$&$(\rm{ppm})$&$(\rm{ppm})$\\
\hline
826.66&0.88&$+0.18$/$-0.15$&9.10&$+4.27$/$-2.91$&3.75&$\pm1.32$\\
882.77&0.89&$+0.16$/$-0.14$&18.56&$+5.30$/$-4.12$&5.18&$\pm1.08$\\
939.58&0.78&$+0.13$/$-0.11$&26.28&$+6.91$/$-5.47$&5.74&$\pm1.02$\\
997.14&0.86&$+0.14$/$-0.12$&42.21&$+9.71$/$-7.90$&7.56&$\pm0.99$\\
1054.33&1.05&$+0.20$/$-0.16$&15.43&$+4.57$/$-3.53$&5.13&$\pm1.11$\\
1111.51&0.97&$+0.20$/$-0.16$&12.83&$+4.36$/$-3.25$&4.54&$\pm1.16$\\
1170.77&0.91&$+0.20$/$-0.17$&3.77&$+3.26$/$-1.75$&2.97&$\pm1.88$\\
1226.34&0.92&$+0.20$/$-0.17$&5.99&$+3.34$/$-2.14$&3.26&$\pm1.46$\\
\hline
\end{tabular}
\end{table*}

Furthermore, we computed the rms amplitudes of the radial modes, $A_{n0}$, according to \citep[e.g.,][]{Chaplin03}:
\begin{equation}
A_{n0}=\sqrt{\frac{\pi}{2} \Gamma_{n0} H_{n0}} \, .
\end{equation}
Amplitudes are always better constrained than the heights and the linewidths themselves. We were careful enough to compute errors on the amplitudes that took into account the correlations between the fitted parameters $\Gamma_{n0}$ and $H_{n0}$ \citep[e.g.,][]{Appourcorr}. These amplitudes were then scaled to their bolometric equivalent using the bolometric correction derived from the spectral response of the \emph{Kepler} passband \citep{Ballotbol}. Finally, we obtained a maximum bolometric amplitude of $A_{n0,{\rm{bol}}}^{\rm{(max)}}\!=\!7.17 \pm 0.75\:{\rm{ppm}}$ (at $833.90\:{\rm{\mu Hz}}$) for KIC~10273246 (cf.~Table \ref{Freq_Mulder2}), and of $A_{n0,{\rm{bol}}}^{\rm{(max)}}\!=\!7.56 \pm 0.99\:{\rm{ppm}}$ (at $997.14\:{\rm{\mu Hz}}$) for KIC~10920273 (cf.~Table \ref{Freq_Scully2}). These values were computed based on the results returned by the respective \emph{best fit} and were found to be consistent with the values obtained from the other fitters' results. 

\citet{KB95} have suggested an empirical scaling relation to predict the amplitudes of solar-like oscillations that, although extensively used, predicts amplitudes in F-type stars that are higher than actually observed \citep[e.g.,][]{Michel08}. Recently, the same authors have proposed a new scaling relation for the amplitudes which is based on simple physical arguments \citep{KB11}:
\begin{equation}
\label{Abol}
A_{\rm{bol}} \propto \frac{L \, \tau_{\rm{mode}}^{0.5}}{M^{1.5}\,T_{\rm{eff}}^{1.25+r}} \, ,
\end{equation}
where $L$ is the stellar luminosity, and the value of $r$ is chosen to be either $r\!=\!1.5$ (assuming adiabatic oscillations) or $r\!=\!2$ \citep[following a fit to observational data in][]{KB95}. By assuming that $\nu_{\rm{max}}$ is a fixed fraction of the acoustic cut-off frequency, i.e.,
\begin{equation}
\nu_{\rm{max}} \propto \nu_{\rm{ac}} \propto \frac{M\,T_{\rm{eff}}^{3.5}}{L} \, ,
\end{equation}
and adopting a scaling relation for the stellar mass based on seismic parameters \cite[e.g.,][]{Kallinger},
\begin{equation}
\label{Mass}
M \propto \Delta\nu^{-4}\,\nu_{\rm{max}}^3\,T_{\rm{eff}}^{1.5} \, ,
\end{equation}
we can combine Eqs.~\ref{Abol}--\ref{Mass} to obtain the following relation (normalized with respect to values in the Sun):
\begin{equation}
\label{Ascaling}
A_{n0,\rm{bol}}^{\rm{(max)}}=A_{n0,\rm{bol}\,\sun}^{\rm{(max)}} \, \left(\frac{T_{\rm{eff}}}{T_{\rm{eff}\,\sun}}\right)^{1.5-r} \, \left(\frac{\nu_{\rm{max}}}{\nu_{\rm{max}\,\sun}}\right)^{-2.5} \, \left(\frac{\Delta\nu}{\Delta\nu_{\sun}}\right)^{2} \, \left(\frac{\Gamma}{\Gamma_{\sun}}\right)^{-0.5} \, , 
\end{equation}
where $A_{n0,\rm{bol}\,\sun}^{\rm{(max)}}\!=\!2.53\:{\rm{ppm}}$, $\nu_{\rm{max}\,\sun}\!=\!3050\:{\rm{\mu Hz}}$, and $\Delta\nu_{\sun}\!=\!135\:{\rm{\mu Hz}}$. Note that by setting $r\!=\!1.5$ (as we will be assuming hereafter), the dependence of Eq.~\ref{Ascaling} on $T_{\rm{eff}}$ is canceled and this scaling relation then solely depends on seismic parameters. Finally, use of this relation gives a predicted maximum bolometric amplitude of $A_{n0,{\rm{bol}}}^{\rm{(max)}}\!=\!7.27 \pm 1.33\:{\rm{ppm}}$ for KIC~10273246, and of $A_{n0,{\rm{bol}}}^{\rm{(max)}}\!=\!8.24 \pm 1.44\:{\rm{ppm}}$ for KIC~10920273, with the relatively large uncertainties dominated by the errors on $\nu_{\rm{max}}$. These values agree with the observed values at the 1-$\sigma$ level. This is a particularly interesting result, since no discrepancy is seen between the predicted and observed amplitudes for the F-type star KIC~10273246.

\subsection{Variation of $\Delta\nu$ with frequency}\label{var_largesep}
An independent methodology has been used to estimate the large
frequency separation and the mode frequencies, which is based on the analysis of the spectrum with the envelope autocorrelation function \citep[EACF;][]{MA09}. As initially proposed by \citet{RV06}, the autocorrelation of the time series -- or, equivalently, the power spectrum of the power spectrum -- windowed with a narrow filter gives the variation of
the large separation with frequency, $\Delta\nu(\nu)$. \citet{Mosser10}
has shown that, with a dedicated comb filter for analyzing the
power spectrum, it is possible to obtain independently the values
of the large separation for the odd ($l\!=\!1$) and even ($l\!=\!0,2$) ridges. We denote them by $\Delta\nu_{\rm{odd}}$ and $\Delta\nu_{\rm{even}}$. Proxies of the mode frequencies can then be integrated from the $\Delta\nu(\nu)$ frequency pattern. In practice, they are derived from the correlation between the observed PDS and a synthetic spectrum based on the $\Delta\nu(\nu)$ pattern.

\begin{figure}[t]
\centering
\includegraphics[width=0.5\textwidth]{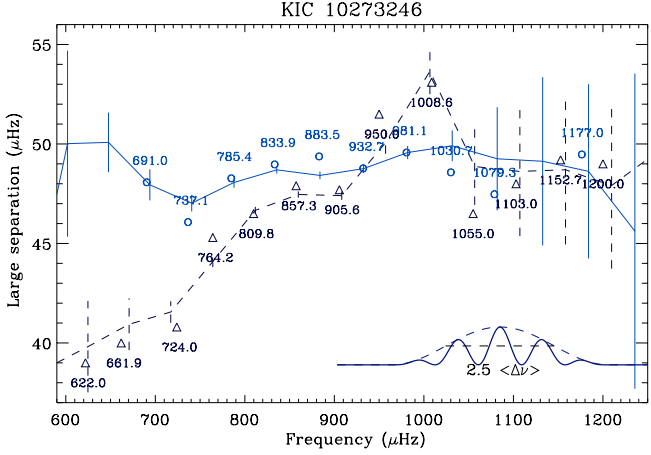}
\caption{Variation of $\Delta\nu_{\rm{odd}}$ and $\Delta\nu_{\rm{even}}$ with frequency for KIC~10273246. The dashed line corresponds to the odd ridge, whereas the solid line corresponds to the even ridge. Error bars are given with the same linestyles. Proxies of the mode frequencies derived with the EACF have been superimposed and their values appended: $l\!=\!0$ (circles) and $l\!=\!1$ (triangles). The inset shows the comb filter used in the analysis.}
\label{fig_autodeltanuridge_10273246}
\end{figure}

The values of $\Delta\nu_{\rm{odd}}$ and $\Delta\nu_{\rm{even}}$ for KIC~10273246 are given in
Fig.~\ref{fig_autodeltanuridge_10273246}. Notice the agreement
between the proxies of the frequencies derived with the EACF and the frequencies of the
\emph{maximal frequency set} (Table \ref{Freq_Mulder}). The EACF emphasizes the
low values of $\Delta\nu_{\rm{odd}}$ as well as its large gradient at the low-frequency end. Also clear is the large discrepancy relative to a regular \'echelle
spectrum around $1000\:{\rm{\mu Hz}}$. The only mode present in the \emph{maximal
frequency set} and not detected with the EACF is the mixed
mode at $695.75\:{\rm{\mu Hz}}$. This peak appears as supernumerary when compared to the regular
agency of the modes. The EACF makes it possible to derive the
large separation one radial order further than does peak-bagging.

\begin{figure}[t]
\centering
\includegraphics[width=0.5\textwidth]{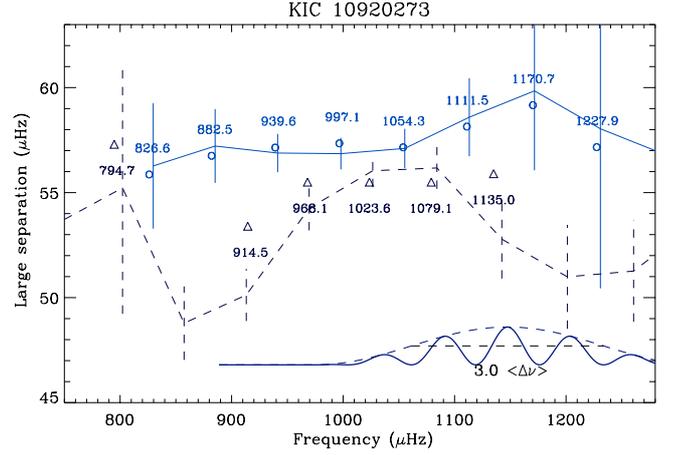}
\caption{Similar to Fig.~\ref{fig_autodeltanuridge_10273246} but for the case of KIC~10920273.} 
\label{fig_autodeltanuridge_10920273}
\end{figure}

Results for KIC~10920273 are given in
Fig.~\ref{fig_autodeltanuridge_10920273}. The lower SNR is
counterbalanced by using a broader filter when computing the EACF. Again,
the analysis is not conclusive for one mixed mode at low frequency,
but it is able to recover the $l\!=\!1$ ridge. We note that the even
ridge is affected by the proximity of mixed modes. As opposed to the case of KIC~10273246, the EACF does not provide any further modes.

\subsection{Rotational splitting and inclination}
Stellar rotation removes the $(2l + 1)$-fold degeneracy of the frequencies of non-radial modes, allowing for a direct measurement of the stellar angular velocity averaged over the regions probed by these modes, as conveyed by Eq.~\ref{ledoux}. Using the radii and masses computed from model-grid-based methods by Creevey et al. (in preparation) together with the estimates of $P_{\rm{rot}}$, we have computed the ratio of the surface angular velocity to the Keplerian break-up velocity, i.e., $\Omega/\sqrt{GM/R^3}$, which returned a value of approximately 1\% for both stars, indicating that these are most likely slow rotators. In view of this and given the precision achievable from the spectra, we have thus decided not to include any second-order effects on the rotational splitting.

The overall profile of a non-radial multiplet thus consists of the sum of $2l + 1$ Lorentzian profiles regularly spaced in frequency, and scaled in height according to the $\mathscr{E}_{l m}(i)$ factors \citep[][]{Dz77,DzG85,GS03}:     
\begin{equation}
	\label{GS}
	\mathscr{E}_{l m}(i) = \frac{(l-|m|)!}{(l+|m|)!} \left[P_l^{|m|}(\cos i)\right]^2 \, ,
\end{equation}
where $i$ is the inclination angle between the direction of the stellar rotation axis and the line of sight, and $P_l^m(x)$ are the associated Legendre functions. Note that $\sum_m \mathscr{E}_{l m}(i) \! = \! 1$, meaning that the $\mathscr{E}_{l m}(i)$ factors represent the relative power contained in the modes within a multiplet. 

\begin{figure}[t]
   \centering
   \includegraphics[width=0.5\textwidth]{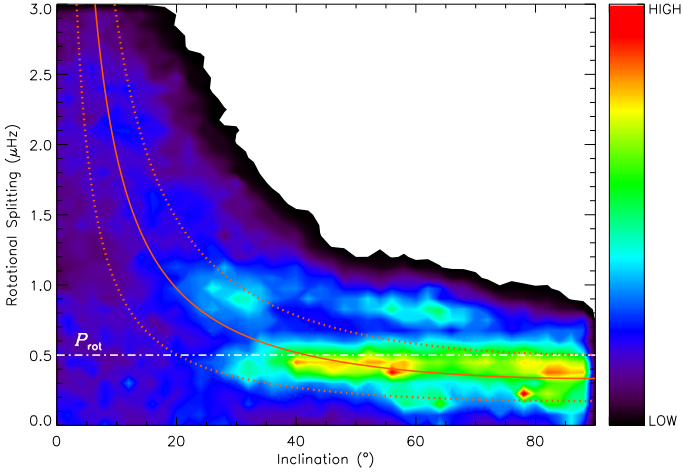}
   \caption{Two-dimensional posterior probability distribution of $\nu_{\rm{rot}}$ and $i$ for KIC~10273246 (the colorbar decodes the probability density level). The prior on $i$ is uniform over the plotted range, whereas the prior on $\nu_{\rm{rot}}$ is uniform over the interval 0--$2\:{\rm{\mu Hz}}$ with a decaying Gaussian wing for higher values of the splitting. The dot-dashed line marks the estimate of the stellar rotational period given in Sect.~\ref{rotmod}. Furthermore, the $P_{\rm{rot}}(i)$ relation (solid curve) of Creevey et al.~(in preparation) is shown together with the corresponding confidence interval (dotted curves).}
   \label{inc_splt_M}
\end{figure}

\begin{figure}[t]
   \centering
   \includegraphics[width=0.5\textwidth]{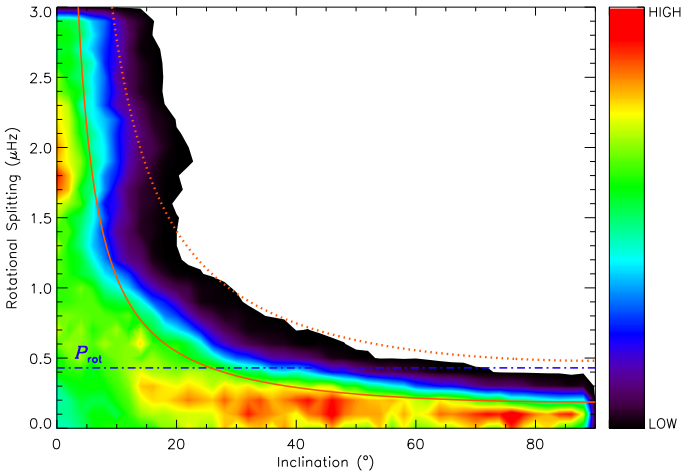}
   \caption{Similar to Fig.~\ref{inc_splt_M} but for the case of KIC~10920273. Given the possibility that $v \sin i\!\approx\!0\:{\rm{km\,s^{-1}}}$, the lower bound of the confidence interval of $P_{\rm{rot}}(i)$ will coincide with the horizontal axis.}
   \label{inc_splt_S}
\end{figure}

While we are not able to robustly constrain the rotational splitting and inclination for both stars, we are however in a position to impose loose constraints on these parameters. Figures \ref{inc_splt_M} and \ref{inc_splt_S} map the two-dimensional posterior probability distributions of these parameters respectively for KIC~10273246 and KIC~10920273, based on the samples from a MCMC analysis of the ten-month-long time series by IAS\_OB. We have overlaid each of these correlation maps with curves representing the estimate of $P_{\rm{rot}}$ given in Sect.~\ref{rotmod} and the $P_{\rm{rot}}(i)$ relation of Creevey et al.~(in preparation), obtained by combining the projected rotational velocity ($v \sin i$) with the stellar radius.

For KIC~10273246, a comparison of the estimate of the rotational period with the $P_{\rm{rot}}(i)$ relation implies that $i\!\gtrsim\!20^{\circ}$, which is also corroborated by the underlying correlation map. Moreover, we notice that the marginal posterior probability distribution of $\nu_{\rm{rot}}$ is unimodal and points toward an interior rotating roughly as fast as the surface. We are however cautious not to claim to have robustly constrained the rotational splitting, since changes in the model -- as well as in the priors on its parameters -- used to represent the background signal may lead to significant alterations in the correlation map.

For KIC~10920273, no constraints on $i$ are possible from a comparison of the estimate of the rotational period with the $P_{\rm{rot}}(i)$ relation. We should nevertheless remind the reader that the estimate of $P_{\rm{rot}}$ is somewhat doubtful (see Sect.~\ref{rotmod}). Although not completely apparent from Fig.~\ref{inc_splt_S}, the marginal posterior probability distribution of $i$ favors the scenario of a star seen pole-on (i.e., $i\!\approx\!0^{\circ}$). Consequently, according to Eq.~\ref{GS}, this would make inviable an inference of the rotational splitting. Going back to the discussion in Sect.~\ref{rotmod}, the possibility that this star is seen pole-on gains strength in explaining the low-SNR peaks at the low-frequency end of the PDS. There is however an alternative interpretation of the correlation map shown in Fig.~\ref{inc_splt_S} that should not be neglected: the high sensitivity of the parameter $i$ to realization noise may lead to what is known as fit locking at $0^{\circ}$ \citep[e.g.,][]{GS03,Ballot08}.

\section{Summary and conclusions}\label{conclusions}

The evolved Sun-like stars KIC~10273246 and KIC~10920273, respectively of spectral types F and G, were observed with \emph{Kepler} at short cadence for approximately ten months (from Q0 to Q4) with a duty cycle in excess of 90\%. The light curves used in the present analysis have been corrected for instrumental effects in a manner independent of the \emph{Kepler} Science Pipeline. Both stars are relatively faint and display low SNR in the p-mode peaks.

Different fitting strategies have been employed to extract estimates of p-mode frequencies as well as of other individual mode parameters, from which we have selected frequency lists that will help constraining stellar models. A total of 30 and 21 modes of degree $l\!=\!0,1,2$ have been identified for KIC~10273246 and KIC~10920273, respectively. These span at least eight radial orders. Furthermore, two avoided crossings ($l\!=\!1$ ridge) have been identified for KIC~10273246, whereas one avoided crossing plus another likely one have been identified for KIC~10920273. These avoided crossings yield strong constraints -- although model-dependent -- on stellar age. Such results confirm previous expectations that asteroseismology of solar-type KASC survey targets is possible down to apparent magnitudes of 11 and fainter, provided we work with a multi-month time series \citep[e.g.,][]{Stello1}.

The peak-bagging results presented in Sect.~\ref{modef} are based on the analysis of data from the first seven months of observations. A smaller number of individual fitters also analysed the ten-month-long time series (from Q0 to Q4). In spite of realization noise being expected to scale as $1/\sqrt{T}$ \citep{Libbrecht,MLE,gapped} thus making it possible to increase measurement precisions, there was barely any gain in terms of the number of modes detected. This is actually not surprising as, once a mode is resolved, the background-to-signal ratio in the power spectrum cannot be improved with time \citep[e.g.,][]{Chaplin03}. The only exceptions would be (i) a radial mode at $690.34\!\pm\!0.23\:{\rm{\mu Hz}}$ for KIC~10273246, which is, however, detected with the EACF and shown in Fig.~\ref{fig_autodeltanuridge_10273246}, and (ii) a $l\!=\!2$ mode at $818.73\!\pm\!0.23\:{\rm{\mu Hz}}$ for KIC~10920273.

Good agreement is found between the observed and predicted mode amplitudes for the F-type star KIC~10273246, based on the revised scaling relation of \citet{KB11}. This is a particularly interesting result that calls for further tests of this scaling relation using the large sample of \emph{Kepler} stars.

Despite blending of the multiplet components of non-radial modes, i.e., $\langle\Gamma\rangle\!\gtrsim\!\nu_{\rm{rot}}$ \citep[e.g.,][]{Ballot06}, we believe to be possible to impose loose constraints on the rotational splitting and stellar inclination for both stars. These constraints are based on a combined analysis involving correlation maps, $P_{\rm{rot}}(i)$ curves and estimates of $P_{\rm{rot}}$. 

The results presented here point towards KIC~10273246 and KIC~10920273 being most likely evolved main-sequence stars. The global asteroseismic parameters reported for these stars, together with a detailed atmospheric analysis, should allow constraining their radius, mass and age with considerable precision (Creevey et al., in preparation). Further insight into the physics of these evolved solar-type stars -- based on detailed modeling and inversion techniques -- is now possible due to the high quality of the seismic parameters found.

\begin{acknowledgements}
Funding for this mission is provided by NASA's Science Mission Directorate. The authors wish to thank the entire \emph{Kepler} team, without whom these results would not be possible. TLC is supported by grant with reference number SFRH/BD/36240/2007 from FCT/MCTES, Portugal. This work was supported by the project PTDC/CTE-AST/098754/2008 funded by FCT/MCTES, Portugal.
\end{acknowledgements}

\bibliographystyle{aa} % style aa.bst 
\bibliography{biblio} % your references

\listofobjects

\appendix
\section{Implementing Peirce's criterion}\label{Peirceappend}
Peirce's criterion is an exact rule for the rejection of doubtful observations derived from the fundamental principles of Probability theory. Quoting \citet{Peirce1}: ``The proposed observations should be rejected when the probability of the system of errors obtained by retaining them is less than that of the system of errors obtained by their rejection multiplied by the probability of making so many, and no more, abnormal observations.''

Logic calls for an iterative assessment of the rejection when one or more observations are rejected. The iteration stops when no further improvement is possible. Based on the work of \citet{Peirce2}, we have implemented Peirce's criterion as follows:
\begin{enumerate}
\item Compute the mean, $\bar x$, and the standard deviation, $\sigma$, for the observational sample $\{x_i\}$;
\item Compute the rejection factor $r$ from \citet{Peirce2} assuming one doubtful observation;
\item Reject observations satisfying $|x_i-\bar x|\!>\!r\sigma$;
\item If $n$ observations are rejected then compute a new rejection factor $r$ assuming $n\!+\!1$ doubtful observations;
\item Repeat steps 3 to 4 until the number of rejected observations no longer increases.
\end{enumerate}

\end{document}